\documentclass{llncs}
\usepackage{graphicx}
\usepackage{balance}  % for  \balance command ON LAST PAGE  (only there!)
\usepackage{array}
\usepackage{float}
\usepackage{color}
\usepackage{url}
\usepackage{threeparttable}
\usepackage{comment}
\usepackage{ulem}

\begin{document}

\title{A Frequency Scaling based Performance Indicator Framework for Big Data Systems}

\author{Chen Yang\inst{1}, Zhihui Du\inst{2}, Xiaofeng Meng\inst{1}, Yongjie Du\inst{1} and Zhiqiang Duan\inst{1}}
\institute{School of Information, Renmin University, China
\and Department of Computer Science and Technology, Tsinghua University, China}
%\author{}
%\institute{}
\maketitle

\begin{abstract}
   It is important for big data systems to identify their performance bottleneck. However, the popular indicators such as resource utilizations, are often misleading and incomparable with each other. In this paper, a novel indicator framework which can directly compare the impact of different indicators with each other is proposed to identify and analyze the performance bottleneck efficiently. A methodology which can construct the indicator from the performance change with the CPU frequency scaling is described. Spark is used as an example of a big data system and two typical SQL benchmarks are used as the workloads to evaluate the proposed method. Experimental results show that the proposed method is accurate compared with the resource utilization method and easy to implement compared with some white-box method. Meanwhile, the analysis with our indicators lead to some interesting findings and valuable performance optimization suggestions for big data systems.
   %In addition, in-memory analytics does not only slow down the Spark due to the memory impact up by $2\times$ but also increase the network impact by $3\times$.
\end{abstract}

\section{Introduction}
Big data systems for large-scale data processing are now in widespread use. To improve their performance, both academia and industry have expended a great deal of effort in identifying their performance bottleneck. The more time a specified resource is used to execute a workload, the larger impact it can change the total performance, and vice versa. The resource with the highest impact (the longest time consumption) is the bottleneck. How to evaluate the resource impact on performance is an essential work to design and tune the big data systems.

Most big data systems use Mapreduce-like frameworks, such as Apache Hadoop and Apache Spark. They allow distributed computing \cite{Dean2004MapReduce} across clusters and always parallelize the use of four major system resources, including CPU, main memory (memory for short), disk and network. It is complex to directly measure the time consumed on different major resources for big data systems. Many researchers\cite{Shi2015Clash,conley2015achieving,Ousterhout2015Making,Wang2017Automating,rightsizingservice} use resource utilizations as indicators to evaluate the resource impact. Many measurement tools have also been developed to monitor the different utilizations\cite{dai2011hitune,Massie2004The}.
However, picking the resource having the greatest utilization as the bottleneck is often not correct and misleading. Different resource utilizations are incomparable with each other, due to different means. For example, the CPU utilization measures percentage of the CPU usage time and the disk bandwidth utilization measures the percentage of the used bandwidth. They are not based on the same metric.

Some researches \cite{Ousterhout2015Making} measure the time consumed on the specified resource directly by adding the fine-gained instrumentations into systems, currently only for the disk and network. However, such \underline{white-box} approaches are too detailed and complicated to implement them easily. In addition, the results are also not accurate. Based on our experiments, the time consumed on I/O resources may be underestimated by $1.6\times$ (seeing Section \ref{section:NSDI15} for details), causing the bottleneck to be misidentified.

Some comparable metrics which can locate the performance bottleneck in an easy way are necessary for big data systems. Unfortunately, the existing approaches cannot work well. We propose a  comparable analysis method to handle this problem instead of the utilization or white-box method. Employing the CPU frequency scaling performance results, our approach can separate the impact of different resources and construct corresponding indicators derived from the same metric. So the value of our indicators are comparable and it is easy to analyze the performance bottleneck based on the proposed indicator framework. The major contributions of this paper are as follows.
 \begin{itemize}
   \item A methodology is proposed to capture the degree of performance impact by measuring how the performance is close to linear speedup when improving the CPU frequency.

   \item Based on the proposed methodology, a comparable performance indicator framework as a \underline{black-box} approach to quantify the impacts of four major resources on big data systems is built. %The framework can separate the impact of different resources and the impact degree of different sources can be ranked by comparing the values of different indicators.

   \item The proposed framework has been employed on a typical Spark based big data system to evaluate its accuracy and efficiency. Furthermore, many interesting findings are gained and many valuable performance optimization suggestions are proposed to  help users tune big data systems like Spark.
 \end{itemize}

The rest of the paper is organized as follows. Section \ref{section:relatedwork} describes the related work. Section \ref{section:resourceDecoupling} presents our approach. Section \ref{section:ExperimentalPreparation} describes the experimental method. Section \ref{section:Experiment} presents our experimental results along with a detailed analysis. Section \ref{section:discussion} discusses how to use our indicator framework efficiently. Section \ref{section:summary} summaries our work and presents directions for future work.

\section{RELATED WORK}\label{section:relatedwork}
The existing researches have extensively studied on the resource impact on the system performance in various flavors, including (1) hardware event counters, (2) resource utilization and (3) resource score. However, they do not provide an easy and comparable approach to evaluate the impacts of four major resources of big data systems.

\textbf{Hardware event counters}. The current computer system provides lots of performance event counters from hardware layer. Hardware events are excellent at capturing how a given piece of hardware is used. Many tools can collect them to help users analyze the performance, such as Perf\cite{Perf} in Linux core and Dtrace\cite{cantrill2004dynamic}, etc. Although they can dynamically trace the system runtime with a low overhead, they never provide the analytical approaches which can quantify the resource bottleneck. In addition, many works focus on interpreting these event counts to analysis the performance bottleneck, including experimental approaches\cite{Sridharan2014Profiling,Sambasivan2011Diagnosing} and modeling approaches\cite{gao2007long,Yoo2012ADP}. However, they currently focus on the low-level indicators. Our approach can generate high-level indicators.

\textbf{Resource utilization}. For profiling a given program to find the bottleneck of major resources, a basic idea is to use resource utilizations. Lots of works \cite{Shi2015Clash,conley2015achieving,Ousterhout2015Making} consider a bottleneck resource with a high resource usage and vice versa. Others\cite{Wang2017Automating,rightsizingservice} simply optimize resource allocation on the basis of resource utilization. However, an important misleading is that resource utilizations are incomparable with each other. The highest utilization might not mean the bottleneck. It might lead to incorrect conclusions of the above works. For example, the blocked time analysis method \cite{Ousterhout2015Making} considers that CPU is the bottleneck resource of Spark by a high CPU utilization in its experiments. Actually, we find that it ignores the memory impact because the classic CPU utilization contains the memory stall cycle, seeing Section \ref{scetion:CRIExperiment} for details.

\textbf{Resource score}. For keeping the comparability, many works give each resource a score and use the score as the resource impact. MIA\cite{yu2018mia} uses the stochastic gradient boosted regression tree to assign the existing indicators the new scores. The new scores are to measure the importance of existing indicators. However, many existing indicators, such as resource utilization, may be not strongly related to the resource impact, so that it cannot correctly find the bottleneck resource. The main goal of our new indicators is to find the bottleneck. Another approach is to run the elaborate benchmark and give a specified resource a unique score to represent the resource performance, such as Spec score for CPU\cite{Spec}. \cite{Dittrich2010Runtime} uses the similar scores to compare the performance of the same resource on different cloud instance types. However, those methods can only evaluate the physics performance of given resources, not the resource impact on a given system.

%The blocked time analysis method \cite{Ousterhout2015Making} for Spark eliminates the disk and network I/O to obtain two maximum speedups if disk or network was infinitely fast. The resource decoupling approach presented by this paper also belongs to this branch, and quantifies the intensiveness degrees of four major components by dynamic decoupling.

\section{Methodology for Building a Comparable Performance Indicator Framework}\label{section:resourceDecoupling}
In this section, we will propose our methodology on how to build a comparable performance indicator framework to identify the performance bottleneck of a big data system.
\subsection{Problem Formulization}
 In this paper, we focus on analyze the resource impact on the end-to-end performance of big data systems using Mapreduce-like framework as the processing engine. We assume that the systems run on a homogeneous cluster with a given resource provisioning and data size. In addition, the system parameters about the resource allocation are also fixed. These requirements mean that we only concern the resource impact on the system under the given configuration. In addition, the load is equally divided among all the tasks, just most of the Spark systems done.

For a given cluster, four major resources are formalized as a vector $R_b=<c_b,m_b,d_b,n_b>$ as the base resource scheme, which represents the CPU including the on-chip cache hierarchy, main memory, disk and network, respectively. Noting that when the CPU is given, we specify $c_b$ as the CPU frequency. With the improvement of CPU frequency, we assume that the memory performance is little or no change and the performance of on-chip cache hierarchy is linear correlation in our cluster. This hypothesis is widely accepted for most x86\_64 computes\cite{Hackenberg2012Memory}. Thus, for a full CPU-intensive workload (i.e., only using the CPU), the performance should change  linearly with the CPU frequency scaling. %If a cluster does not meet these hypotheses cited above, our method may not work.

Cited above, our methodology is to observe the performance improvement with the CPU frequency scaling. We first define the performance improvement. For a given cluster, $CF=\{c_{1}, c_{2},...,c_{l}\}$ is the CPU frequency set from the same CPU where $c_{j} \ge c_{i}$ if $j \ge i$ and $c_{1} \ge c_{b}$. We can easily scale the CPU frequency on modern CPUs. $DB=\{d_{1},d_{2},...,d_{m}\}$ is the disk set, where the performance of $\forall {d_j} \in DB$ is better than $d_b$. $NB=\{n_{1}, n_{2},...,n_{z}\}$ is the network bandwidth set, where the performance of $\forall {n_k} \in NB$ is faster than $n_b$. The set about memory has not been defined due to not upgrade memory.

$RT(c, d, n)$ is the running time of one workload, where the resource scheme $<c,m_b,d,n>$ is configured to a cluster. When the CPU frequency goes up to $c_i \in CF$ from $c_b$ and the other resources are fixed, we can define the CPU performance improvement degree $CPI$ as

\begin{equation}\label{eq:CPUSpeedup}%\nonumber
  CPI(c_i,d,n)=1-\frac{RT(c_i,d,n)}{RT(c_{b},d,n)},
\end{equation}
where $CPI\in[0,1)$. If it is closer to 1, the performance improvement is higher.
\begin{figure*}[tbp]
\centering
\includegraphics[width=5in,height=1.5in]{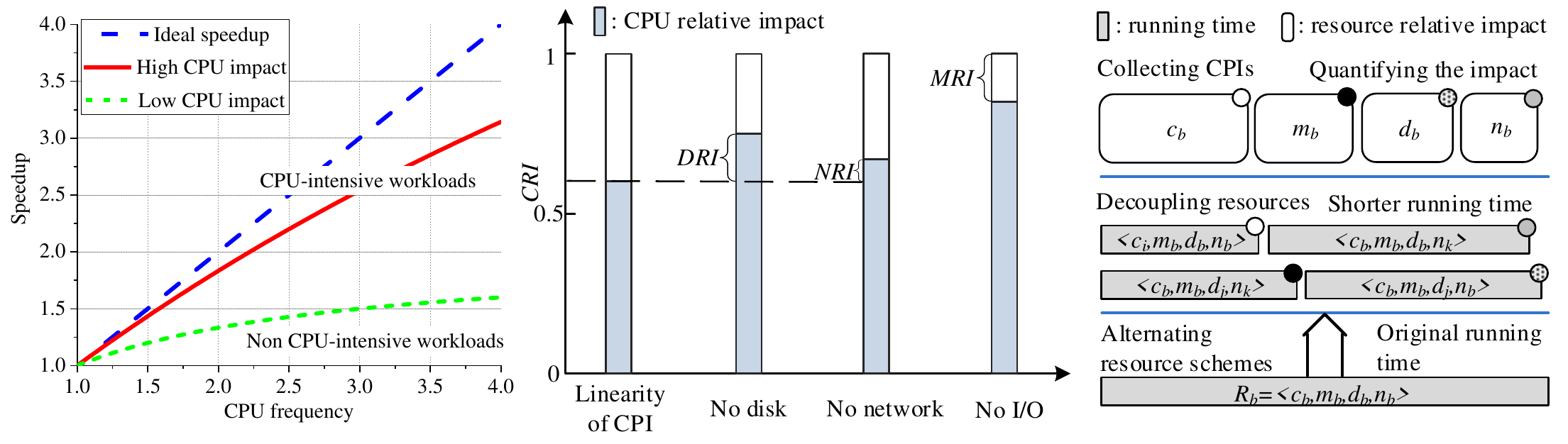}
    \caption{Left: The speedup (i.e., $RT(c_{b},d,n)/RT(c_i,d,n), c_{b}=1.0$) on different kinds of workloads when improving the CPU frequency. Middle: The impacts of non-CPU resources are derived from the variation of CPU impact. Right: We decouple the resource impacts by alternating resource schemes and each resource scheme corresponds to an evaluation of resource impact.}
\label{fig:RecourceDecoupling}
\end{figure*}
\subsection{Performance Indicator Definition}\label{section:Indicator}
For a given big data system, the execution time can be formalized as follows.

\begin{equation}\label{eq:mapreducePerformance}
RT = {\theta _1}\frac{{scale}}{machine} + {\theta _2}\log \left( machine \right) + {\theta _3}machine + {\theta _4},
\end{equation}
where $scale$ is the data size and $machine$ is the cluster size\cite{venkataraman2016ernest}. The first item is the computation time, including CPU impact and memory impact, i.e., the time consumed on the CPU and the memory. The rest are communication time and fixed cost, mainly including disk impact and network impact, i.e., the time consumed on the disk and the network.

Cited above, both data size and cluster size are fixed in our scenario. By improving CPU frequencies, we can only reduce the CPU impact of Eq. (\ref{eq:mapreducePerformance}), causing the performance improvement. In this way, we can demonstrate the relation between the system performance and CPU frequency. For easy understanding, we use the speedup, not $CPI$ in Figure \ref{fig:RecourceDecoupling} but their features are similar. For a CPU-intensive system, it will be always on-CPU and rarely be blocked by I/O or memory stalls. The CPU is the only limiting resource. Therefore, the speedup is linearly proportional to the improvement of the CPU frequency in the ideal case. Obviously, if the CPU impact is low at Eq. (\ref{eq:mapreducePerformance}), CPU frequency scaling will have little impact on the performance (i.e, low speedup), showing the high non-CPU impact. It motivates us to understand both CPU impact and non-CPU impact by observing the non-linear change in performance.

\textbf{CPU Relative Impact}. With the improvement of CPU frequency, we can define the linearity of performance improvement as CPU relative impact ($CRI$) to correlate the CPU impact. For assigning $CRI \in [0,1]$, we define $1-c_{b}/c_i$ ($c_i \in CF$) as the upper bound of the performance improvement. If $CPI(c_i,d,n)$, instead of speedup, is closer to $1-c_{b}/c_i$, systems are more CPU-intensive. If $CPI(c_i,d,n)$ is closer to 0, systems are not CPU-intensive. To describe this relationship, we formalize $CRI$ on $R_b$ as
\begin{equation}\label{eq:cb}
CRI(R_b) = \frac{1}{l}\sum\limits_{{c_i} \in CF} {\frac{{CPI({c_i},d,n)}}{{1 - {c_{b}}/{c_i}}}},
\end{equation}
where $l=|CF|$ is the number of alternative CPU frequencies, and $CRI\in[0, 1]$. For $\forall{c_i \in CF}$, if $CPI({c_i},d,n)=1-c_{b}/c_i$, then $ CRI=1$, the workloads will be full CPU-intensive. On the other extreme, for $\forall{c_i \in CF}$, if $CPI(c_i,d,n)=0 $ then $CRI=0$, the CPU has no impact on the system performance.

\textbf{Disk Relative Impact}. If the disk is upgraded to $d_j \in DB$, the upper bound of the performance improvement being similar to $1-c_{b}/c_i$ is unknown, so we cannot use a method similar to Eq. (\ref{eq:cb}) to evaluate the disk relative impact ($DRI$). Actually, if we could eliminate the disk blocked time, the system will tend to more CPU-intensive leading that $CRI$ will be higher. It in essence correlates the disk impact to the change of $CRI$ by the CPU frequency scaling. Thus, we can identify the disk impact from the change of $CRI$. We can eliminate the disk blocked time by upgrading the disk in Figure \ref{fig:RecourceDecoupling} and use the increment of $CRI$ to define $DRI$ as
\begin{equation}\label{eq:db}
DRI(R_b) =\mathop {\max} \limits_{{{d}_{j}} \in DB} (CRI\left( {c_b,m_b,d_j,n_b} \right) - CRI\left( {R_b} \right)),
\end{equation}
where $m=|DB|$ is the number of alternative disks, and $DRI\in[0, 1]$. If $DRI \to 0$, the disk has no impact on the system performance. On the other extreme, if $DRI \to 1$, the system is full disk-intensive. In addition, the upgraded disks may introduce different performance improvements due to sequential and random access. However, the precision of $DRI$ is dependent on the performance of upgraded disk, so that the equation suggests that the optional disk should maximize $CRI$, otherwise the evaluated $DRI$ will be small.

\textbf{Network Relative Impact}. The same method for the disk can be used to evaluate the network relative impact ($NRI$) as
\begin{equation}\label{eq:nb}
NRI(R_b) = \mathop {\max} \limits_{{{{n}}_{k}} \in NB} (CRI\left( {c_b,m_b,d_b,n_k} \right) - CRI\left( {R_b} \right)),
\end{equation}
where $z=|NB|$ is the number of alternative networks, and $NRI\in[0, 1]$. This is similar to $DRI$, where $NRI \to 1$ represents highly network-intensive systems and vice versa.

\textbf{Memory Relative Impact}. Because the performance of different consumer memories are so close to each other, we cannot identify the memory impact by upgrading the memory hardware. For example, our test finds that STREAM \cite{stream} (an intensive memory access benchmark) with DDR3-1600 RAM is only 4.2\% faster than with DDR3-1333 RAM. Thus, the performance improvement is hard to be observed by using the faster memory. From another perspective, we can eliminate the I/O impact (disk and network) as much as possible, leading the system to be only impacted by the CPU and memory. Based on this observation, we define memory relative impact ($MRI$) as
\begin{equation}\label{eq:mb}
MRI(R_b) = 1 - \mathop {\max}\limits_{{{d}_{j}} \in DB, {{{n}}_{k}} \in NB}(CRI\left( {c_b,m_b,d_j,n_k} \right)),
\end{equation}
where $MRI\in[0, 1]$. A high $MRI$ means a memory-intensive big data system.

\section{Experimental Method}\label{section:ExperimentalPreparation}
In this section, we use our approach to analyze Spark's performance, including two running modes, which have different performance characteristics, so that they can be considered as two systems. The detailed cluster setup and running mode are as follows.
\begin{figure}[t]
\centering
\includegraphics[width=3.2in,height=2.2in]{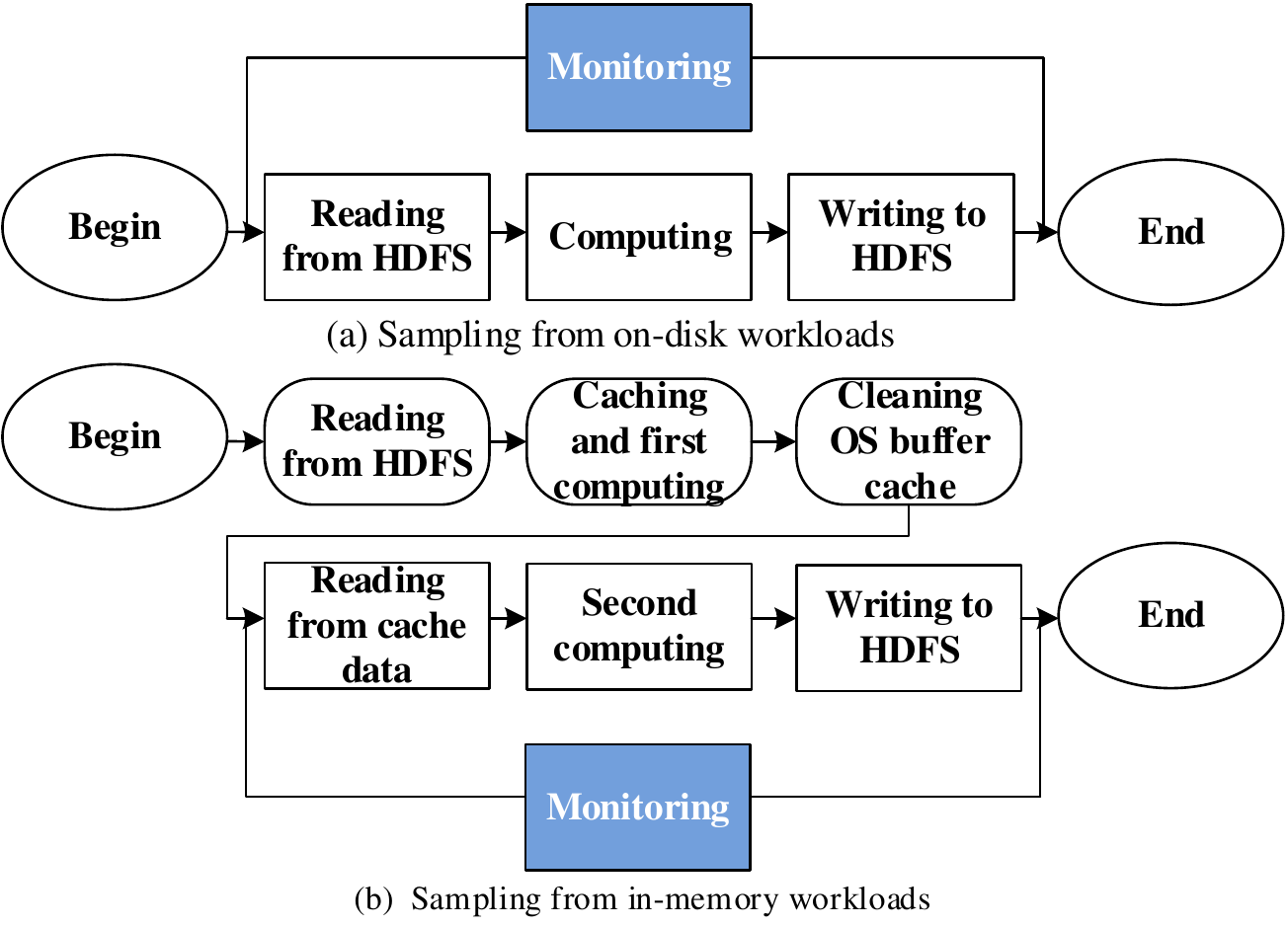}
\caption{The procedures of disk mode and memory mode}
\label{fig:DiskandMemoryMode}
\end{figure}
\subsection{Cluster Setup}
In our experiment, the baseline resource scheme is set as $R_b=\{1.2GHz,DDR3\textrm{-}$\\$1600,HDD,1Gbps\}$ and other resource sets are $CF=\{2.4GHz,3.6GHz\}$, $DB=\{SSD\}$ and $NB=\{5Gbps,10Gbps\}$. The processor is Intel i7-4790, which has 8 logical CPU cores. We investigate the long-term scientific data services of our collaborators and find that the CPU frequency is usually set at a low level to save energy. Therefore, we set the base CPU frequency to 1.2GHz to get close to the production environment. For disk, we upgrade the disk by replacing HDD with SSD that has the same capacity. The performance of SSD can eliminate most of disk blocked time for our experimental setup. In addition, the software environment, cluster configuration, and data distribution are identical between HDD and SSD. For network, we have a fiber-optic 10 Gbps network environment. Here, 1 Gbps and 5 Gbps can be obtained by the network speed limit using \texttt{tc}. Our cluster has 10 nodes with 1 master and 9 slaves, where every node has 1 processor, 32GB RAM and two 500GB SATA disks.

We build our system environment by using stable versions of the software (Apache Spark 1.6.3, Apache Hadoop 2.6.0, and 64-bit Ubuntu 14.04 Server), where Hadoop and Spark use the same machine as the master node. Our approach can also analyze the performance of Spark 2.x, but more interesting findings are got from Spark 1.6.3, so that we do not present the results of Spark 2.x. We run Spark on the standalone and create one worker configured with 8 threads and 28GB RAM (i.e., one thread per CPU). We store the input data in HDFS (Hadoop Distributed File System), and the output data will also be written to HDFS.

 %In addition, we set the JVM heap size to 2GB for the namenode process and 1 GB for the datanode process. For Linux, the maximum socket connections are increased to 2048, and we also set the number of files that a process may have opened at a time to 65535 to prevent HDFS errors.
We use two SQL benchmarks, i.e., BDBench\cite{Pavlo2009A} (50GB Gzip data) and TPC-DS\cite{Nambiar2006The} (40GB Parquet data\cite{Parquet}). BDBench has 9 queries and in TPC-DS we choose 42 queries. The chosen benchmarks are comprehensive and can cover most of the basic operations for big data.

\subsection{Running Mode}\label{section:disk_and_memory_mode}
Every query is finally parsed into on-disk and in-memory workloads (i.e., in-memory analytics) by Spark. For disk mode, on-disk workloads read the input data from HDFS and write the output data to HDFS, and we collect the running time over the whole process, as shown in Figure \ref{fig:DiskandMemoryMode}(a). For memory mode, in-memory workloads read input data from the memory and write output data to HDFS. As shown in Figure \ref{fig:DiskandMemoryMode}(b), we have to run every workload twice because the cache function is not a action operator\cite{ApacheSpark}. The first running is to cache data and the second running is just called as the memory mode, which will be monitored. We run both of them, collect the running time and clean the OS buffer cache after the workloads are finished. For each queries, we repeat the whole sampling procedure 3 times and use the average results to reduce the error.

%\begin{figure}[t]
%\centering
%\includegraphics[width=5in,height=1.7in]{bottlenckDstribution}
%\caption{Workload distributions with different bottlenecks on $R_b$. If a workload with $XRI \ge 0.5$, we label it into X-intensive.}
%\label{fig:bottlenckDstribution}
%\end{figure}
\begin{figure}[t]
\centering
\includegraphics[width=3.2in,height=1.5in]{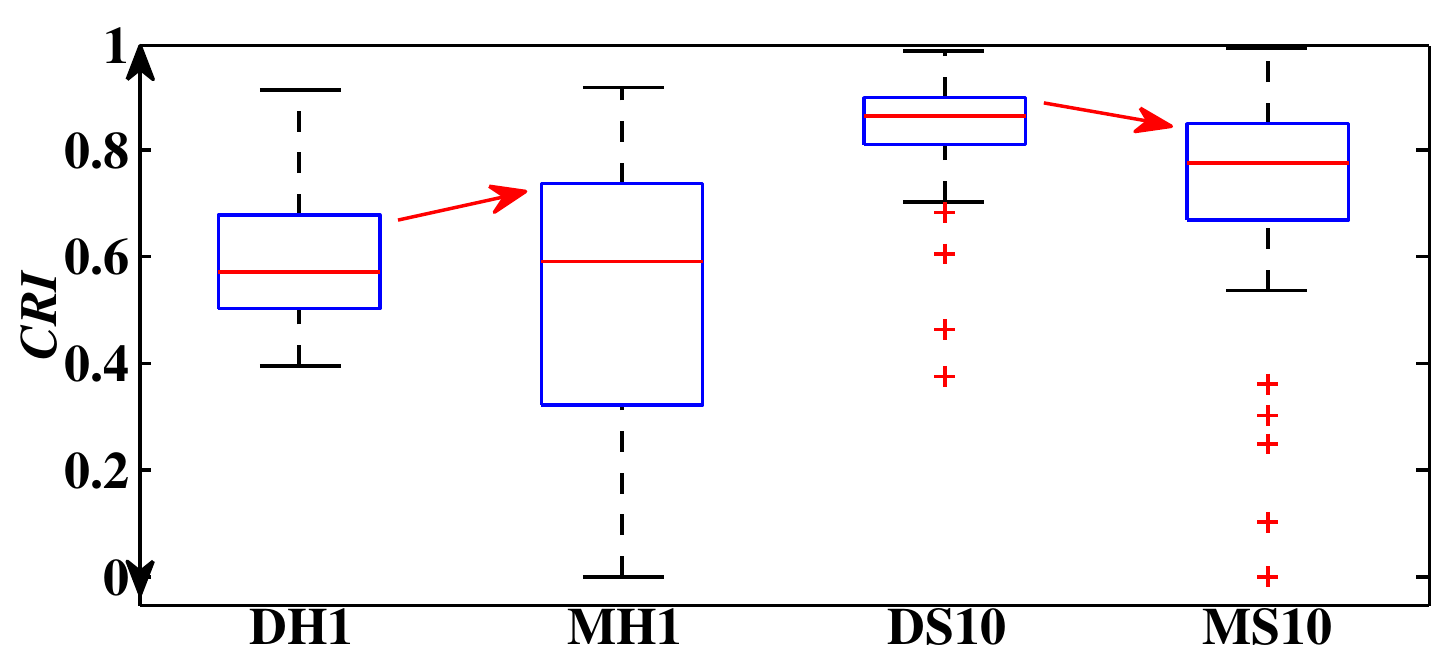}
\caption{CPU relative impact. On the x-axis, ``DH1" and ``MH1" represent disk mode and memory mode on HDD and 1Gbps, respectively. ``DS10" and ``MS10" are disk mode and memory mode on SSD and 10Gbps, respectively.}
\label{fig:CPUbound}
\end{figure}
\begin{table}[tbp]
\caption{Average Resource Relative Impact on $R_b$}
 \centering
\begin{tabular}
{|m{3cm}<{\centering}|m{3cm}<{\centering}|m{1.6cm}<{\centering}|m{1.5cm}<{\centering}|m{1.5cm}<{\centering}|}
\hline
\textbf{Resource relative impact}&
\textbf{Running mode}&
\textbf{BDBench}&
\textbf{TPC-DS}&
\textbf{Avg} \\
\hline
\raisebox{-1.50ex}[0cm][0cm]{$CRI$}&
Disk mode&
0.73&
0.58&
0.61 \\
\cline{2-5}
 &
Memory mode&
0.55&
0.52&
0.53 \\
\hline
\raisebox{-1.50ex}[0cm][0cm]{$MRI$}&
Disk mode&
0.04&
0.18&
0.16 \\
\cline{2-5}
 &
Memory mode&
0.18&
0.31&
0.3 \\
\hline
\raisebox{-1.50ex}[0cm][0cm]{$DRI$}&
Disk mode&
0.17&
0.25&
0.24 \\
\cline{2-5}
 &
Memory mode&
0.19&
0.2&
0.2 \\
\hline
\raisebox{-1.50ex}[0cm][0cm]{$NRI$}&
Disk mode&
0.04&
0.015&
0.02 \\
\cline{2-5}
 &
Memory mode&
0.06&
0.06&
0.06 \\
\hline
\end{tabular}
\label{tab:resourceimpact}
\end{table}

\section{Experimental Results and Analysis}\label{section:Experiment}
In this section, we split the whole experiment into five parts to study the resource impact on Spark. We compare our performance indicator framework with the resource utilization (e.g., CPU utilization, disk bandwidth utilization and network bandwidth utilization) and the time blocked white box analysis method\cite{Ousterhout2015Making}.

\subsection{CPU Impact Analysis}\label{scetion:CRIExperiment}
 Table \ref{tab:resourceimpact} shows $CRI$ solved by Eq. (\ref{eq:cb}), where the label ``Avg" represents the average of both benchmarks. Overall, $CRI$ is, on average, 0.57 for both running modes, where 76\% queries in disk mode are CPU-intensive and 64\% queries in memory mode are CPU-intensive, i.e, $CRI \ge 0.5$. It suggests that the CPU is the bottleneck for Spark. Curiously, for both benchmarks, $CRI$ in memory mode is always lower than it in disk mode. This implies that Spark is more CPU-intensive when reading input data from the disk. We put BDBench and TPC-DS together to find that $CRI$ in memory mode tends to two extreme poles, as shown in ``DH1" and ``MH1" of Figure \ref{fig:CPUbound}.

When $CRI\ge 0.6$, the approximate median 51\% of in-memory workloads are only slightly greater than the 45\% of on-disk workloads. This phenomenon shows the memory mode has more CPU-intensive workloads. However, too many in-memory workloads are low CPU-intensive. When $CRI\le 0.4$, 27\% are in-memory workloads (approximately 87\% of them in TPC-DS), far more than the 2\% of on-disk workloads. This phenomenon that reading data from memory has a lower CPU impact can be be reasonably explained as follows.
\begin{itemize}
  \item Reading the cache data causes a lower LLC (Last Level Cache) hit rate, so the memory stall time gets longer. Details are provided in Section \ref{section:MIDAnalysis}.
  \item The high $CRI$ is incurred when decompressing input data in disk mode, but it is not required for memory mode. Relatively speaking, memory mode will show the low CPU impact. Details are provided in Section \ref{sectiondisk}.
  \item Spark is blocked by the network I/O more frequently in memory mode. This phenomenon can also reduce $CRI$. Details are provided in Section \ref{section:networkBottleneckAnaysis}.

\end{itemize}

  %\textbf{The trend of CPU speedup}. Overall, the current CPU can speed Spark up by average 35\%. Also, Spark is not absolute CPU-bound for two phenomena. In addition, (1) $CPUSpeedup$ is less than the limit speedup, and the gap is widening gradually with the improvement of CPU frequency, showing the CPU speedup is not nearly linearly proportional to the improvement. The reason is that other bottlenecks are not small enough to be ignored, but are more obvious accordingly. (2) With the improvement of CPU frequency, disk mode easily can be speeded up compared with memory mode, but the reason is different for both benchmarks. In BDBench, memory mode is more disk-bound than disk mode, detailed in Section \ref{Sectiondiskspeedup}. In TPC-DS, some workloads on memory mode cannot be speeded up well, detailed in Section \ref{sectionMemory-bound}.
%\begin{figure}[t]
%\centering
%\includegraphics[width=3.5in,height=1.5in]{CPUutil}
%\caption{CPU utilization on HDD and 1Gbpsp}
%\label{fig:CPUutil}
%\end{figure}

\begin{figure}[t]
\centering
\includegraphics[width=5in,height=3in]{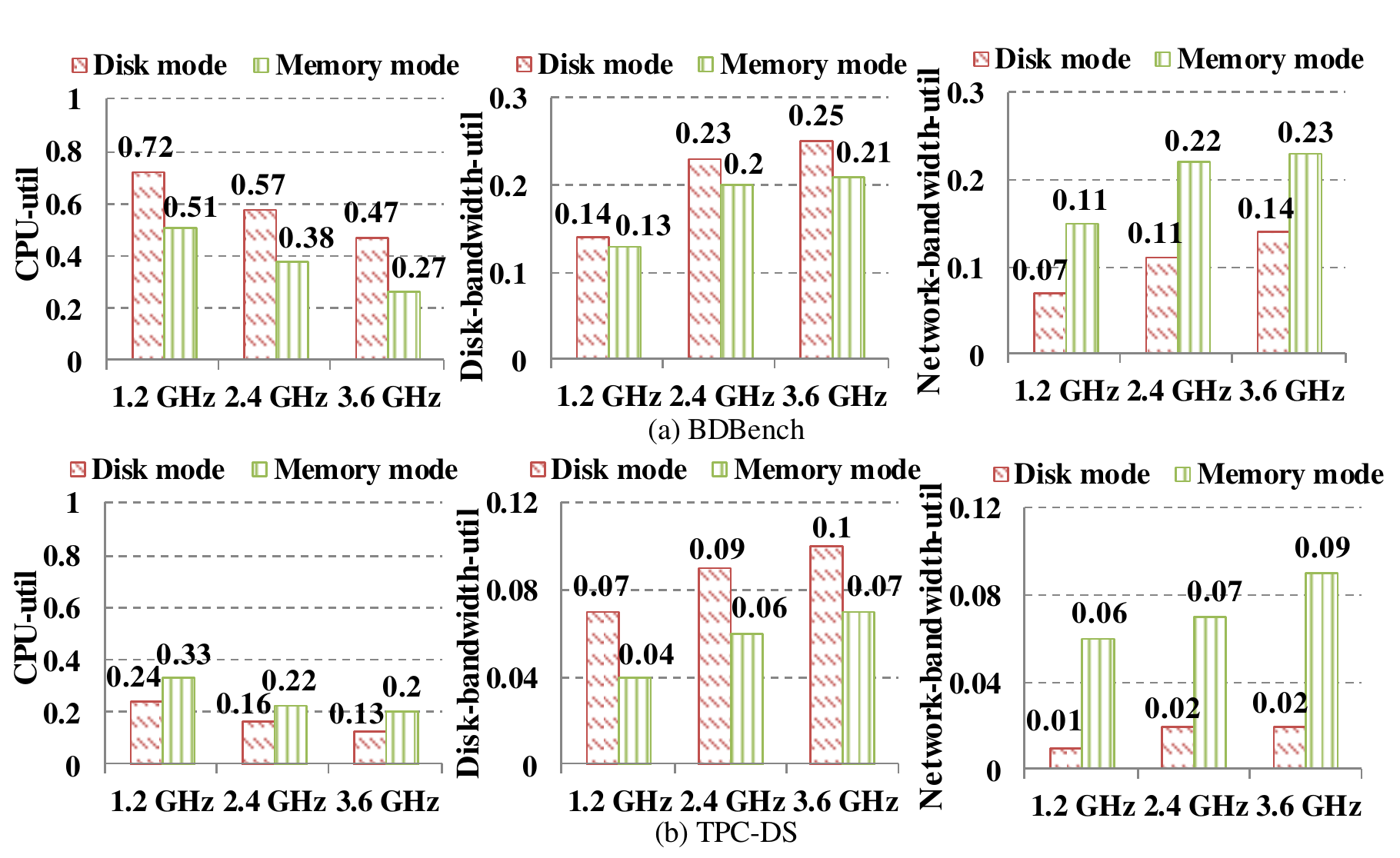}
\caption{Different resource utilizations on HDD and 1Gbps when scaling the CPU frequency}
\label{fig:resourceUtil}
\end{figure}

%\begin{table}[tbp]
%\caption{Average CPU Relative Impact}
% \centering
%\begin{tabular}
%{|m{2.2cm}<{\centering}|m{2.2cm}<{\centering}|m{1.5cm}<{\centering}|m{1.5cm}<{\centering}|m{1.5cm}<{\centering}|}
%\hline
% &\textbf{BDBench}&\textbf{TPC-DS}&\textbf{Avg}\\
%\hline
%Disk mode&0.73&0.58&0.61\\
%\hline
%Memory mode&0.55&0.52&0.53 \\
%\hline
%\end{tabular}
%\label{tab:AverageCID}
%\end{table}

\textbf{Why CPU utilization is not a good representation of the CPU impact on performance?} For most of performance analysis scenarios, the CPU utilization (CPU-util) is usually used to evaluate the CPU impact on performance. A high CPU utilization means that applications are more CPU-intensive and vice versa. Strictly speaking, it is not accurate enough.

In Figure \ref{fig:resourceUtil}, for TPC-DS CPU-util in memory mode is greater than it in disk mode. This trend contradicts $CRI$. The reason is that LLC misses are more frequent causing the high memory stall time in memory mode of TPC-DS and CPU-util includes the memory stall time. Too many stall cycles cause a high CPU-util. However, the memory stall time should be the memory impact, not the CPU impact. This finding suggests that a high CPU-util cannot always represent a high CPU impact.

For both benchmarks, especially in TPC-DS, CPU-util is very low, but $CRI$ suggests that the CPU is the bottleneck. This is also contradictory. The reason is that CPU-util only shows the system impact on the CPU (the CPU usage), not the CPU impact on the system (the percentage of the CPU usage time). For a multicore processor, in most cases, CPU-util is the average utilization of all cores. For Spark, because of scheduling delay and task difference, the scheduler cannot ensure that the task threads always run on all cores. Thus, in our experiments, it is normal that some CPU cores are idle but other CPU cores are always in use, causing the low CPU-util. However, it does not mean that the CPU consumes less time than other resources. This finding suggests that a low CPU-util cannot represent a low CPU impact.

\textbf{Suggestion}. CPU-util and $CRI$ can be combined to help users give the optimized suggestions. For example, the high CPU-util and the low $CRI$ may imply that the system has a weak memory management strategy. The low CPU-util and the high $CRI$ may imply that the CPU cores are not fully used.

\begin{figure}[t]
\centering
\includegraphics[width=3.5in,height=1.7in]{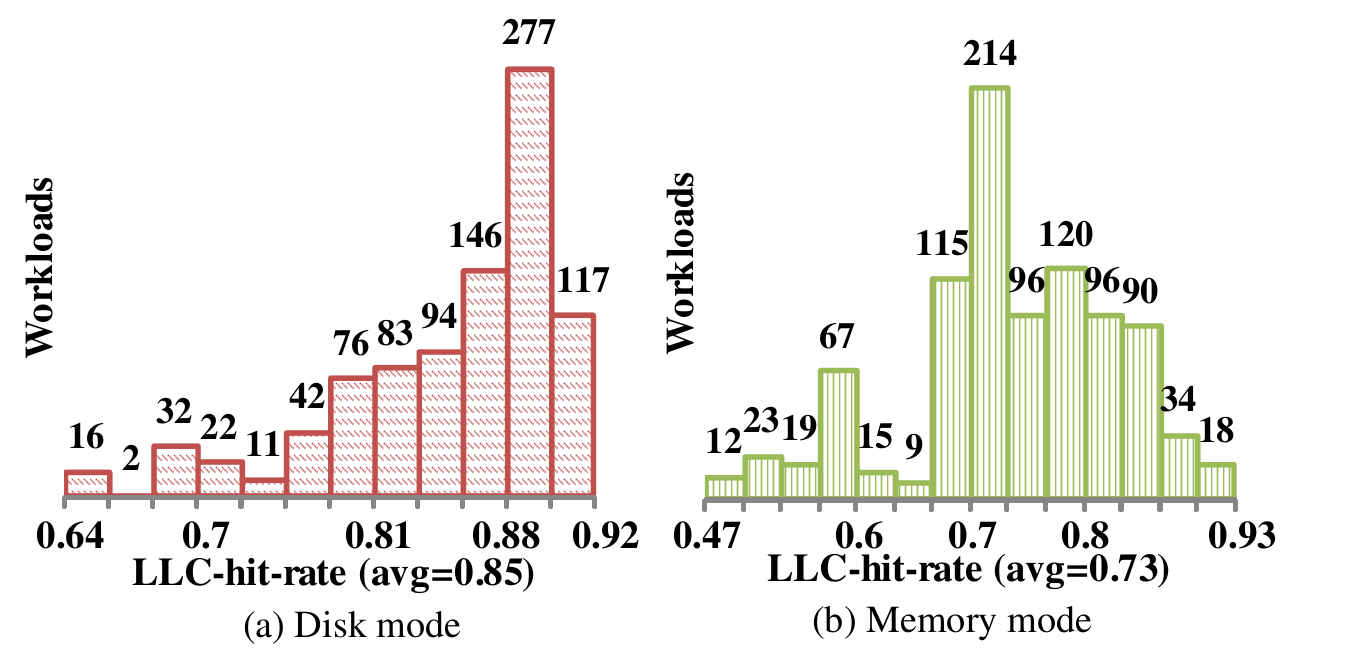}
\caption{Distributions of LLC hit rate on all of resource schemes}
\label{fig:TheDistributionofLLC}
\end{figure}

%\begin{table}[tbpp]
%\caption{Average Memory Relative Impact}
% \centering
%\begin{tabular}
%{|m{2.2cm}<{\centering}|m{1.5cm}<{\centering}|m{1.5cm}<{\centering}|m{1.5cm}<{\centering}|}
%\hline
%&\textbf{BDBench}&\textbf{TPC-DS}&\textbf{Avg}\\
%\hline
%Disk mode&0.04&0.18&0.16\\
%\hline
%Memory mode&0.18&0.31&0.3 \\
%\hline
%\end{tabular}
%\label{tab:AverageMID}
%\end{table}
\subsection{Memory Impact Analysis}\label{section:MIDAnalysis}
As shown in Figure \ref{fig:CPUbound}, we focus on ``DS10" and ``MS10". There is an abnormal phenomenon. After we only upgrade the I/O resources, $CRI$ in memory mode shows a downward trend relative to $CRI$ in disk mode, compared with the upward trend on HDD and 1Gbps. It reveals that the memory impact in memory mode is $2\textrm{-}4.5\times$  greater than it in disk mode. Based on this, we determine $MRI$ by Eq. (\ref{eq:mb}), as shown in Table \ref{tab:resourceimpact}. Overall, the average $MRI$ is 0.23. For both benchmarks, $MRI$ in memory mode is always greater than it in disk mode. This shows that reading the cache data makes Spark more memory-intensive.

 \textbf{Why is memory mode more memory-intensive?} It is worth to note the LLC hit rate. In Figure \ref{fig:TheDistributionofLLC}, we demonstrate the distribution of LLC hit rate, where the interval of every bar is nearly equal. For memory mode, the average and highest bars have 14-21\% deteriorations compared with those in disk mode. This shows that the performance of LLC hit in memory mode is weak.

Especially when we run TPC-DS on SSD and 1 Gbps, memory mode (Running time is average 56.7s) is unexpectedly slower than disk mode (average 55.7s). The high overhead of moving data into the CPU even exceeds the advantage of caching data in memory, causing caching data to have no effect. Therefore, for memory mode reading cache data causes $MRI$ to be increased.

The cache operation for Spark 1.6.3 is important because it is the basis for in-memory data analytics. In Spark SQL, the main idea of the cache strategy for structured data is as follows. When data are cached in memory, Spark stores them in a two-dimensional array into a columnar format. This data structure is conducive to in-memory compression. However, when the data is processed, Spark must transform them from columnar format into a row format. This transformation can break the data locality, leading to a reduction of the LLC hit rate. For Spark 2.x, we also find the similar phenomenon.

\textbf{Suggestion}. The high $MRI$ implies that the performance of Spark SQL in memory mode can be improved by optimizing the cache operation. Actually, we find that reading cache data in memory mode has to transform a columnar array into a row array, causing the performance reduction.

\subsection{Disk Impact Analysis}\label{sectiondisk}
\begin{figure}[tbp]
\centering
\includegraphics[width=4.3in,height=1.6in]{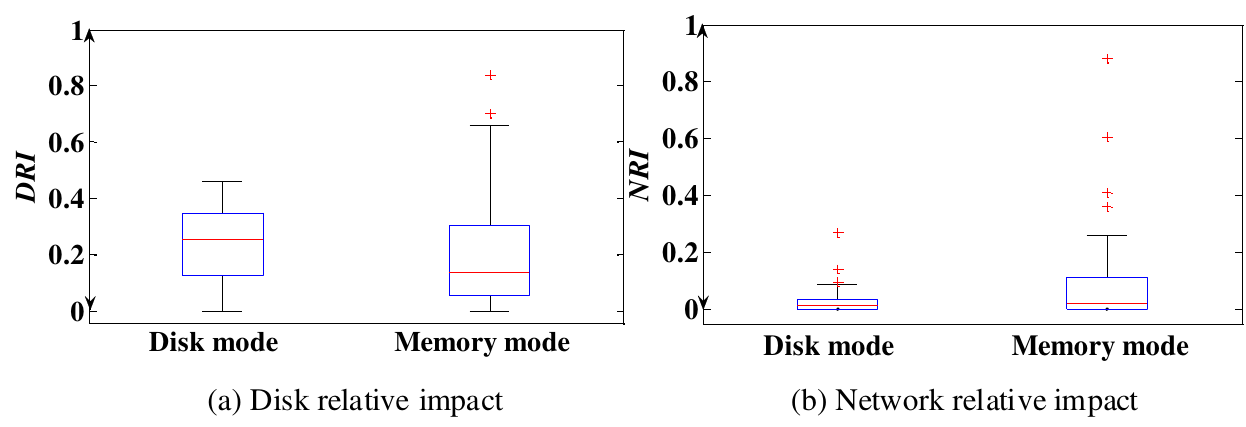}
\caption{Disk relative impact and network relative impact}
\label{fig:diskAndNetworkBound}
\end{figure}

We solve $DRI$ using Eq. (\ref{eq:db}), and the average is 0.22 for both modes. Disk mode is more disk-intensive than memory mode, as shown in Table \ref{tab:resourceimpact}. This trend is also demonstrated in Figure \ref{fig:diskAndNetworkBound}(a). However, for different benchmarks, the trend is the opposite. For BDBench, memory mode is more disk-intensive. For TPC-DS, disk mode is more disk-intensive. Especially for BDBench, the abnormal trend of $DRI$ suggests that reading input data from disk does not necessarily represent that the system is more disk-intensive. We also show the traditional indictor (i.e., disk bandwidth utilization) in Figure \ref{fig:resourceUtil}. It suggests that disk mode is more disk-intensive for BDBench, being different from the trend of $DRI$. Actually, we think the disk bandwidth utilization cannot accurately reflect the disk impact.
%\begin{figure}[tbp]
%\centering
%\includegraphics[width=3.5in,height=1.5in]{diskBandwidthUtil}
%\caption{Disk bandwidth utilization on HDD and 1Gbps}
%\label{fig:diskBandwidthUtil}
%\end{figure}

\textbf{What factors cause the difference of DRI?} Combined with our experiments, we summarize two reasons causing different disk impacts for both benchmarks as follows.

\textbf{Compression for BDBench}. Memory mode is more disk-intensive in BDBench. The additional decompression in disk mode increases much computation, causing the relatively low disk impact. The only difference between disk mode and memory mode is reading compression data from the disk or not. $CRI$ in disk mode is 0.73, far more than 0.55 in memory mode, showing that many CPU cycles are used for the decompression. Relatively speaking, the disk impact is low in disk mode, compared with memory mode. Therefore, the high disk bandwidth utilization does not mean the high disk impact.

\textbf{Short tasks for TPC-DS}. Disk mode is more disk-intensive in TPC-DS. Unlike the long-calculation tasks in BDBench, the short tasks in TPC-DS easily are blocked by I/O due to the less overlap. The current big data systems leverage the asynchronous I/O mechanism to overlap the computation and I/O request to improve the performance. However, in disk mode of TPC-DS the CPU overhead is relatively low for each task, causing the overlap to be disabled. Thus, the disk impact increases. For example, the low CPU-util in TPC-DS is on average 17\% in disk mode and 25\% in memory mode, showing the lower opportunity for overlap. In addition, TPC-DS has many short tasks (e.g., 0.7 second per task in TPC-DS, compared with 9.1 seconds per task in BDBench) also showing the less overlap. It leads to an interesting phenomenon that $DRI$ is excellent in TPC-DS, but the disk-bandwidth-util is very low, as shown in Figure \ref{fig:resourceUtil}. Therefore, the low disk bandwidth utilization does not mean the low disk impact.

\textbf{Suggestion}. The high disk-bandwidth-util and the low $DRI$ imply that the system has a good I/O performance with less disk blocked time when reading much data. In contrast, the low disk-bandwidth-util and the high $DRI$ mean a weak I/O performance.
The SQL's optimizer should build the long computing tasks or merge I/O requests as much as possible to maximize the overlap.

\subsection{Network Impact Analysis}\label{section:networkBottleneckAnaysis}
As shown in Table \ref{tab:resourceimpact}, we solve $NRI$ using Eq. (\ref{eq:nb}), and the average is 0.04 for both modes, which is minimal within the four major resources. $NRI$ in memory mode is $1.5\textrm{-}4\times$ greater than it in disk mode in Figure \ref{fig:diskAndNetworkBound}(b). The network bandwidth utilization in Figure \ref{fig:resourceUtil} also has the same trend.

\textbf{Why does memory mode have a higher NRI?} Spark uses the network in the following three stages. (1) The input data may be read from remote disks, but this rarely happens (ie.g., only 5\% of the data from remote disks in our cluster), because HDFS preferentially reads input data from local disks. (2) The shuffle read stage needs to read data from both local and remote disks. (3) Writing output data to HDFS needs to backup two duplicates to remote disks. Because HDFS rarely reads input data from remote disks, the network I/O has nothing to do with the decompression. Moreover, shuffle and output stages are the same in both modes. Thus, the amount of data transferred over the network is nearly equal in both modes (The difference is, on average, 5.6\% in our cluster). Due to the shorter running time in memory mode, more data are transferred over the network per second. This is manifested as a higher network impact.

\textbf{Suggestion}. Memory mode needs to transfer data over the network more frequently, causing the higher $NRI$. Combined with the result that BDBench's memory mode have a higher disk impact than disk mode in the previous section, it is actually necessary for users to pay more attention to the I/O impact, rather than the CPU impact for in-memory analytics in some cases.

\subsection{Inaccuracy of Time Blocked White-Box Method}\label{section:NSDI15}
The blocked time analysis method \cite{Ousterhout2015Making} for Spark is used for analyzing the impacts of the disk and network. It collects the I/O blocked time by adding some instrumentations into the system and simplifies part of shuffle I/O into the upper bound of the disk I/O or network I/O. Finally, simulate the infinitely fast disk or network by ignoring I/O blocked time to evaluate Spark's maximum performance improvement. Actually, it mainly evaluates the I/O impact, i.e., both disk and network.

This method relies on adding some instrumentations into HDFS's core to get the blocked time when Spark accesses HDFS. However, the corresponding codes have not been opened, so that only shuffle I/O can be profiled\cite{trace-analysis}. Even so, we design several cases to illustrate the limitations of this approach.

\textbf{Major page faults}. Spark is usually impacted by the external factors, such as OS. The intra-system instrumentations cannot monitor them. For example, the system execution is not only blocked when reading data from disk, may also be blocked due to major page faults issued by OS. We design a simple experiment to demonstrate this problem. In BDBench, q3C is the most complex query. When we use 56 GB compressed data to run Spark with q3C in our cluster, Spark will be starved for memory. We run q3C without output in memory mode on $R_b$. For contrast, this query also runs when I/O resources are upgraded to SSD and 10Gbps. Time blocked analysis method shows that q3C can be sped up by 48.6\% ($ < 50\%$), suggesting that q3C might not be the I/O bottleneck, but it is actually sped up by 77.7\%, suggesting that q3C is definitely the I/O bottleneck. The I/O impact is underestimated by $1.6\times$. Actually, Spark on HDD is slowed down due to major page faults (6,394 per node) but it can be significantly sped up by SSD. This phenomenon is ubiquitous. In 42 queries of TPC-DS, approximately 79\% of them have major page faults. Overall, the I/O impact might be underestimated.

Compared with the blocked time analysis method, our indictor framework does not only focus on the impacts of four major resources, but it can also evaluate the latent I/O impact by upgrading I/O resources.

\section{Discussion}\label{section:discussion}
In this section, we discuss the major characteristics of our indicator framework and how to use our method efficiently.

\textbf{Comparability}. Our indicator framework is built from $CRI$, so their values are comparable. Thus, the greatest one can be identified as the bottleneck. It is noted that the sum of them is not necessarily equal to 1. When we upgrade the disk and network simultaneously for calculating $MRI$, the improvement of $CRI$ may be not equal to the sum of $DRI$ and $NRI$ by upgrading the disk or network separately.

\textbf{Scalability}. The resource replacement method may limit our indicators on large-scale clusters. Our indicators are only dependent on the end-to-end performance in essence. Thus, we can leverage the performance prediction technique to achieve the scalability. For example, Ernest\cite{venkataraman2016ernest} can predict the end-to-end performance of the large-scale MapReduce-like workloads by training a performance model with the performance data from different small-scale clusters. Thus, we can run the system on small-scale clusters with our indicator framework and train the performance model by Ernest. Further, we can predict the resource impact on large-scale clusters.

\textbf{Cost}. For our indicator framework, the major cost is upgrading the disk and network. However, it is not necessary to upgrade I/O resources for CPU-intensive applications, e.g., $CRI>0.5$. This observation is very helpful to reduce the usage cost of our indicators.

\textbf{Accuracy}. The workload and the alternative resources can affect the accuracy of the value of our indicators. If the load can be equally divided among all the tasks, our indicator framework will identify the performance change as accurate as possible when upgrading resources. To evaluate the linearity of performance improvement with $CRI$, we run the system at different $c_i \in CF$ and use the average as the $CRI$ to improve the accuracy. For three other indicators, it is easily to achieve the upgrade of I/O resources. For the disk, both SSD and main memory are used as $d_j$. The RamDisk technique\cite{koutoupis2009linux} supported by Linux core can use the main memory as the disk. For the network, both fiber-optic 10Gbps network or faster InfiniBand network architecture can be used as $n_k$.

\section{Summary \& Future Work} \label{section:summary}
In this paper, we propose a performance indicator framework to evaluate the relative impact of four major resources on big data systems. Values of different indicators can be built based on measuring the CPU frequency scaling performance results. Many experiments are done to verify our approach and Spark's performance is analyzed in depth. We summary the most important advantages of our framework. In addition, many interesting findings are found and some valuable suggestions are given to help users tune their Spark.

\textbf{Advantages}. First, our four indicators are \emph{comparable} with each other because they are derived from the same metric. The feature ensures the bottleneck can be easily found through our approach. Second, our indictors are more \emph{accurate} compared with the resource utilization and the existing white-box approach because our approach is strongly related to the time consumed on the specified resource. Therefore, our approach can easily find the underlying performance issues.  Third, our approach is also \emph{easy to implement} relying on the general CPU frequency scaling technology.

\textbf{Findings}. (1) The CPU impact may go down when Spark reads data from memory instead of disk, because lower LLC hit rate, no data decompress, and more frequent network blocking will happen. (2) Using CPU utilization as CPU performance impact indicator is often misleading because long memory stall time may lead to high CPU utilization and unbalanced tasks/threads scheduling on multicore systems will lead to low CPU utilization. (3) Reading data from memory will significantly increase the memory impact by $2\textrm{-}4.5\times$ because lower LLC hit rate will happen. Sometimes the performance will be lower than reading data from disk because data locality is broken. (4) Disk bandwidth utilization is also often misleading to identify the disk impact because it cannot show how much disk time can be overlapped with CPU time. (5) The network impact is often the lowest for most Spark big data systems. But its value  can be increased by $1.5\textrm{-}4\times$ when Spark reads data from memory.

\textbf{Suggestions}. Even though resource utilizations are often misleading, with the help of our indicator framework, it is easy for us to not only find the cause, but also give the method to handle the problem. So we can combine the two methods together to identify some Spark's potential problems on the memory management strategy, the scheduler and the SQL's optimizer. Some specific tuning suggestions can also be given from our work. For example, users should pay more attention to the impact of I/O resources when executing in-memory analytics, rather than ignore them.

In the future, we will focus on the absolute resource impact, such as direct evaluating the time consumed by different major resources. It will be a more powerful tool which does not only identify the bottleneck but also predict the potential performance gains when optimizing some resource.

\section{Acknowledgement}
This research was partially supported by the grants from National Key Research and Development Program of China (No. 2016YFB1000602, 2016YFB1000603); Natural Science Foundation of China (No. 91646203, 61532016, 61532010, 61379050, 61762082); Fundamental Research Funds for the Central Universities, Research Funds of Renmin University (No. 11XNL010); and Science and Technology Opening up Cooperation project of Henan Province (172106000077).
\bibliography{peformance}
\end{document}